\documentclass[pra,twocolumn,showpacs]{revtex4-1}

\usepackage{amsmath}
\usepackage{graphicx}

\newcommand{\bra}[1]{\langle #1|}
\newcommand{\ket}[1]{|#1\rangle}

\begin{document}

\title{Entanglement distillation for continuous-variables under a thermal environment: Effectiveness of a non-Gaussian operation}

\author{Jaehak Lee}
\author{Hyunchul Nha}
\affiliation{Department of Physics, Texas A \& M University at Qatar, P.O. Box 23874, Doha, Qatar}

\begin{abstract}
We study the task of distilling entanglement by a coherent superposition operation $t\hat{a}+r\hat{a}^\dagger$ applied to a continuous-variable state under a thermal noise. In particular, we compare the performances of two different strategies, i.e., the non-Gaussian operation $t\hat{a}+r\hat{a}^\dagger$ is applied before or after the noisy Gaussian channel. This is closely related to a fundamental problem of whether Gaussian or non-Gaussian entanglement can be more robust under a noisy channel and also provides a useful insight into the practical implementation of entanglement distribution for a long-distance quantum communication. 
We specifically look into two entanglement characteristics, the logarithmic negativity as a measure of entanglement and the teleportation fidelity as a usefulness of entanglement, for each distilled state. We find that the non-Gaussian operation after (before) the thermal noise becomes more effective in the low (high) temperature regime.
\end{abstract}

\maketitle

\section{\label{sec:introduction}Introduction}

In quantum information processing, it is an important task to distribute entanglement between distant parties, but entanglement can be easily degraded due to interaction with a noisy environment. Numerous schemes to overcome the decoherence by using nondeterministic local operations were proposed to distill entanglement for discrete variable systems \cite{PhysRevLett.76.722,PhysRevA.54.3824,PhysRevLett.77.2818}. In the continuous variable (CV) regime, it is known that Gaussian states cannot be distilled by using only Gaussian operations \cite{PhysRevLett.89.137903,PhysRevLett.89.137904,PhysRevA.66.032316}. A frequently used CV entangled resource, i.e. two-mode squeezed vacuum (TMSV), is a Gaussian state and a certain non-Gaussian operation is thus required. Distillation of Gaussian states was initially studied with two elementary non-Gaussian operations, namely, a single photon subtraction $\hat{a}$ and addition $\hat{a}^\dagger$  \cite{PhysRevA.61.032302,PhysRevA.65.062306,PhysRevA.67.032314,PhysRevA.73.042310}. The entanglement distillation by the photon subtraction scheme was experimentally realized \cite{takahashi2010entanglement}. Recently, it was also found that a more efficient distillation can be achieved by a coherent superposition of photon subtraction and addition, $t\hat{a}+r\hat{a}^\dagger$ \cite{PhysRevA.84.012302}.

These studies for CVs, however, did not take into account the interaction with a noisy environment that actually motivated the topic of entanglement distillation. They have simply shown that the non-Gaussian operations can distill a pure Gaussian entangled state into a pure non-Gaussian entangled state with higher entanglement. The task of entanglement distillation is more important when the state becomes mixed due to interaction with a noisy environment. Only a few works so far treated this practically important problem, which found that distillation by a single photon subtraction is possible even in the presence of noise \cite{PhysRevA.82.062316}.

In this paper, we investigate how efficiently we can distill CV entanglement under a thermal environment by a coherently superposed operation $t\hat{a}+r\hat{a}^\dagger$. While the previous analysis for the distillation of mixed CV entangled state considered the case of applying the operation after entangled states are distributed through noisy channels  \cite{PhysRevA.82.062316}, we also study another strategy. That is, a non-Gaussian operations is applied before a CV entangled state is distributed to distant parties. Which of the two strategies gives a better performance is closely related to a fundamental problem of to what extent Gaussian and non-Gaussian entangled states can be robust under a noisy channel. Some evidences were put forward to support the conjecture that Gaussian entanglement is more robust than non-Gaussian entanglement \cite{PhysRevLett.105.100503,PhysRevA.83.024301}. However, there exist some counterexamples in which non-Gaussian states can be more robust than Gaussian states \cite{PhysRevLett.107.130501,PhysRevLett.107.238901,NJP}. From a practical point of view, it is crucial to have a longer survival time of entanglement, as entanglement cannot be distilled at all once it dies out.

We consider a TMSV (a prototype of CV entangled state) as an initial state and apply $t\hat{a}+r\hat{a}^\dagger$ (an elementary non-Gaussian operation that includes the photon subtraction and the addition as special cases)  before or after the noisy channel [Fig. \ref{fig:strategy}].
We will show that two different strategies have advantages in different temperature regimes. A non-Gaussian entangled state distilled before the thermal noise can survive longer than a Gaussian entangled state without any operation. This effect is particularly remarkable in the high-temperature regime where the survival time of Gaussian entanglement is short. A Gaussian state that becomes separable via a noisy channel cannot be distilled into an entangled state because local operations and classical communication (LOCC) cannot create any entanglement from a separable state. However, if the entanglement survives in a Gaussian state, it can be distilled into a highly entangled state by the coherent superposition operation. In the low-temperature regime where Gaussian entanglement survives long enough, it turns out that the non-Gaussian operation after the noisy channel enhances entanglement better.

This paper is organized as follows. In Section \ref{sec:description}, we briefly introduce the description of two-mode entangled states and their tranformations by a coherent superposition operation and under a thermal noisy channel. Then, we study the entanglement properties of output states, namely, the logarithmic negativity in Section \ref{sec:negativity} and the teleportation fidelity in Section \ref{sec:fidelity}. We summarize our results and discuss the applicability to practical protocols in Section \ref{sec:conclusion}.

\section{\label{sec:description}CV entangled states and noisy channel}

\subsection{Two-mode squeezed vacuum and coherent superposition operation}

We start with a Gaussian entangled state generated by two-mode squeezing, which is described by
\begin{equation}
\ket{\Psi}_\mathrm{TMSV}= \mathrm{sech} s \sum_{n=0}^{\infty} \tanh^n s \ket{n,n} ,
\end{equation}
where $s$ is squeezing parameter and $\ket{n,m}$ represents a two-mode state in Fock state basis. The corresponding characteristic function is given by

\begin{eqnarray}
\chi_\mathrm{TMSV}(\xi_1,\xi_2) & = & \mathrm{Tr} \left[ \ket{\Psi}_\mathrm{TMSV}\bra{\Psi}_\mathrm{TMSV} \hat{D}_1(\xi_1)\hat{D}_2(\xi_2) \right] \nonumber \\
& = & \exp \left[ -\frac{1}{2} \left( |\xi_1|^2 + |\xi_2|^2 \right) \cosh 2s \right. \nonumber \\
& & \qquad \left. - \frac{1}{2} \left( \xi_1 \xi_2 + \xi_1^* \xi_2^* \right) \sinh 2s \right] ,
\end{eqnarray}
where $\hat{D}_i(\alpha)\equiv e^{\alpha\hat{a}_i^\dagger-\alpha^*\hat{a}_i}$ denotes the displacement operator acting on mode $i=1,2$ with amplitude $\alpha$ \cite{Barnett}.

A coherently superposed operation $t\hat{a}+r\hat{a}^\dagger$ can be implemented by erasing the which-path information of the trigger photon emerging from the photon subtraction $\hat{a}$ or the photon addition $\hat{a}^\dagger$ \cite{PhysRevA.82.053812}. The parameters $t$ and $r$ can be controlled by adjusting the transmitivity of the beamsplitter that erases the which-path information before photodetection. 
When this operation is applied to the mode $i$ of a multi-mode state $\rho$, whose original characteristic function is $ \chi(\overrightarrow{\xi}) = \chi(\xi_1,\xi_2,\cdots) $, the state is transformed to
\begin{eqnarray}
\lefteqn{ \left( t\hat{a}_i+r\hat{a}_i^\dagger \right) \rho \left( t\hat{a}_i^\dagger+r\hat{a}_i \right) } \nonumber \\
& \to & \lefteqn{ O_i(\chi(\overrightarrow{\xi})) } \nonumber \\
& & = \left[ t\left( -\frac{\partial}{\partial\xi_i^*} + \frac{\xi_i}{2} \right) + r\left( \frac{\partial}{\partial\xi_i} - \frac{\xi_i^*}{2} \right) \right] \nonumber \\
& & \qquad \left[ t\left( \frac{\partial}{\partial\xi_i} + \frac{\xi_i^*}{2} \right) + r\left( -\frac{\partial}{\partial\xi_i^*} - \frac{\xi_i}{2} \right) \right] \chi(\overrightarrow{\xi}) .
\end{eqnarray}

\subsection{Noisy Gaussian channel}

On the other hand, when the $i$th mode propagates through a thermal-noise channel, its evolution can be described by a master equation
\begin{equation}
\dot{\rho} = \frac{\Gamma}{2} n_\mathrm{th} L[\hat{a}_i^\dagger] \rho + \frac{\Gamma}{2} (n_\mathrm{th}+1) L[\hat{a}_i] \rho.
\end{equation}
Here $\Gamma$ and $n_\mathrm{th}$ are the loss coefficient and the average photon number in thermal environment, respectively. With current technology, the loss coefficient in optical fiber can be made less than a few $\mathrm{dB/km}$. $L[\hat{A}]$ is the Lindblad operator defined by $ L[\hat{A}]\rho \equiv 2\hat{A}\rho \hat{A}^\dagger - \hat{A}^\dagger\hat{A}\rho - \rho\hat{A}^\dagger\hat{A} $. Here, we assume that both of the two modes propagate under the thermal channels with the same $\Gamma$ and $n_\mathrm{th}$.

Instead of solving the master equation, the interaction with the thermal noise can be described by a beamsplitter model. The input state $\rho$ and the thermal ancilla state $\rho_{\mathrm{th}}=\frac{e^{-\beta \hat{a}^\dagger\hat{a}}}{{\rm Tr}\{e^{-\beta \hat{a}^\dagger\hat{a}}\}}$ with mean photon number $n_\mathrm{th}=\frac{1}{e^\beta-1}$ are mixed at a beam splitter with transmissivity $\eta=e^{-\Gamma t}$ and the ancilla mode is then traced out in the output. That is,
\begin{equation}
\rho^\prime = \mathrm{Tr}_{\rm th} \left[ U_{\mathrm{BS}} \rho \otimes \rho_{\mathrm{th}} U_{\mathrm{BS}}^\dagger \right] ,
\end{equation}
where $\mathrm{Tr}_{\rm th}$ is the partial trace over the ancilla mode and $U_{\mathrm{BS}}=e^{\theta(\hat{a}_i^\dagger\hat{a}_{\rm th}-\hat{a}_i\hat{a}_{\rm th}^\dagger)}$ describes the beamsplitter interaction of $i$th mode and ancilla mode ($\cos\theta=\sqrt{\eta}$). Then, the characteristic function for the output state turns out to be
\begin{eqnarray}
\chi^\prime(\overrightarrow{\xi}) & = & N_i(\chi(\overrightarrow{\xi})) \nonumber \\
& = & \chi_\mathrm{th}(\sqrt{1-\eta}\xi_i) \times \nonumber \\
& & \; \chi(\xi_1,\cdots,\xi_{i-1},\sqrt{\eta}\xi_i,\xi_{i+1},\cdots) ,
\end{eqnarray}
where $ \chi_\mathrm{th}(\xi) = e^{-\frac{1}{2}(2n_\mathrm{th}+1)\xi^2} $ is the characteristic function of a thermal state  with mean photon number $n_\mathrm{th}$.

\begin{figure}[t]
\centering \includegraphics[clip=true, width=0.95\columnwidth]{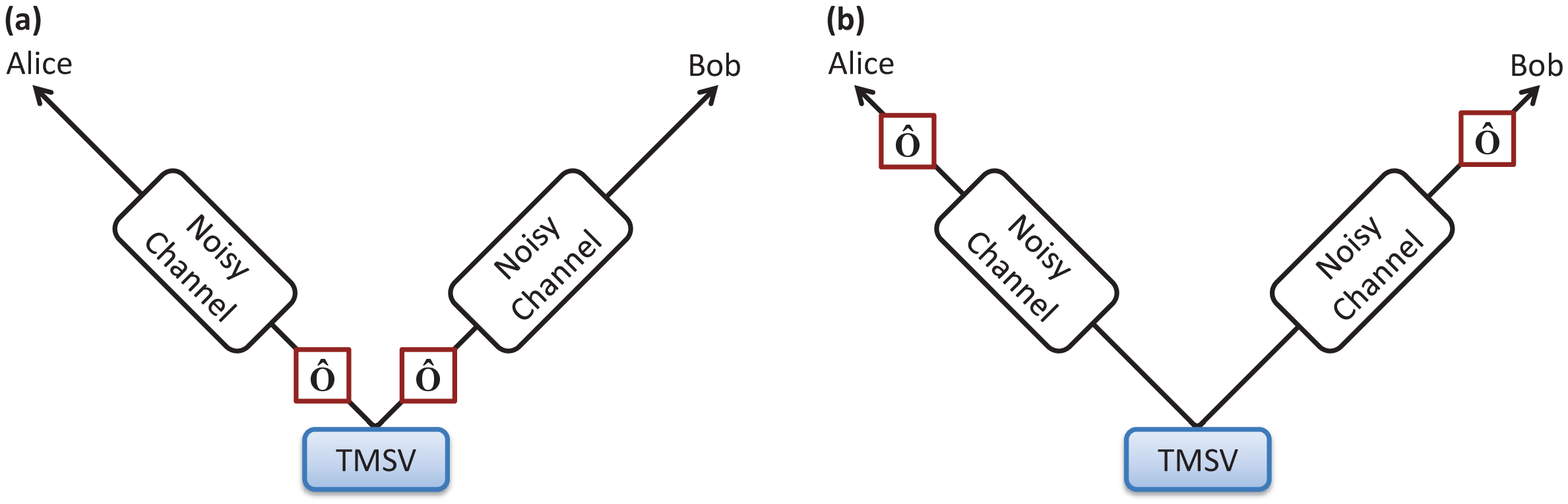}
\caption{\label{fig:strategy}Schematic diagram for CV entanglement distillation under noisy channels}
\end{figure}

Combining Eqs. (3) and (6), one can obtain the characteristic functions of final states via two different ways of distillation [Fig. \ref{fig:strategy}] using the characteristic function of initial state $\chi_\mathrm{TMSV}$ with operators $O_i$ and $N_i$ ($i=1,2$). 
If we perform distillation before sending the initial state through noisy channels [Fig. \ref{fig:strategy} (a)], the output state will be given by
\begin{equation} \label{eq:cfon}
\chi_\mathrm{no}(\xi_1,\xi_2) = N_1 \circ N_2 \big( O_1 \circ O_2 (\chi_\mathrm{TMSV}(\xi_1,\xi_2)) \big) .
\end{equation}
On the other hand, if we perform distillation after sending the state through noisy channels [Fig. \ref{fig:strategy} (b)], the output state will be given by
\begin{equation} \label{eq:cfno}
\chi_\mathrm{on}(\xi_1,\xi_2) = O_1 \circ O_2 \big( N_1 \circ N_2 (\chi_\mathrm{TMSV}(\xi_1,\xi_2)) \big).
\end{equation}
In the following, we compare the entanglement properties under two different strategies with the corresponding characteristic functions $\chi_\mathrm{no}$ and $\chi_\mathrm{on}$.

\section{\label{sec:negativity}Logarithmic Negativity}

In this section, we first quantify the degree of entanglement for each output state under the two different schemes [Fig. \ref{fig:strategy}]. As a measure of entanglement, we particularly adopt the logarithmic negativity, an easily computable entanglement monotone \cite{PhysRevA.65.032314,PhysRevLett.95.090503}.

\subsection{Estimation of Logarithmic Negativity}
The logarithmic negativity is defined as
\begin{equation}
E_N(\rho) \equiv \log_2 ||\rho^{T_A}||_1
\end{equation}
where $\rho^{T_A}$ is the partial transpose of the density matrix $\rho$ and the trace norm is defined as $ ||A||_1 \equiv \mathrm{tr}\sqrt{A^\dagger A} $.

For a Gaussian state, which can be completely described with its first and second momenta of canonical operators $x_i=\frac{1}{\sqrt{2}}(a_i+a_i^\dagger)$ and $p_i=\frac{1}{i\sqrt{2}}(a_i+a_i^\dagger)$, the logarithmic negativity can be easily calculated from its second-moment covariance matrix $\sigma$. The logarithmic negativity of a Gaussian state is given by \cite{PhysRevA.65.032314}
\begin{equation} \label{eq:cmnegativity}
E_N(\sigma) = \mathrm{max}\left\{ 0,-\log_2 (2\tilde{d}_{-}) \right\}.
\end{equation}
where $\tilde{d}_{-}$ represents the least symplectic eigenvalue of partially transposed covariance matrix $\sigma^{T_A}$.

On the other hand, the distillation operation we consider is non-Gaussian, thus Eq. (\ref{eq:cmnegativity}) is not adequate to fully address the logarithmic negativity by incorporating all higher-order momenta. For a non-Gaussian state, we investigate the density matrix elements in the number-state basis. 
For a numerical calculation of negativity, we restrict our consideration to the subspace of finite photon number states. That is, we calculate  the negativity with a truncated density matrix $ \rho_\mathrm{trunc} = \hat{P}_0\rho\hat{P}_0 $ where $ \hat{P}_0 = \sum_{n,m=0}^{N_\mathrm{trunc}} \ket{n,m}\bra{n,m}$ with the truncation number $N_\mathrm{trunc}$. Note that the logarithmic negativity of $\rho_\mathrm{trunc}$ cannot be larger than that of $\rho$ due to the monotonicity of $E_N$. The trace norm $||\rho^{T_A}||_1$ is a monotone under a positive partial transpose preserving operation (PPT operation) which maps $\rho$ into $ \rho_i = \Psi_i(\rho)/\mathrm{tr}\Psi_i(\rho)$ with its probability $p_i=\mathrm{tr}\Psi_i(\rho)$ \cite{PhysRevLett.95.090503}. That is,
\begin{equation}
||\rho^{T_A}||_1 \ge \sum_i ||\Psi_i(\rho^{T_A})||_1 .
\end{equation}
Since the set of operations $ \{ \hat{P}_0, \hat{P}_1=I-\hat{P}_0 \} $ is a PPT operation, we have
\begin{eqnarray}
||\rho^{T_A}||_1 & \ge & ||(\hat{P}_0\rho\hat{P}_0)^{T_A}||_1 + ||(\hat{P}_1\rho\hat{P}_1)^{T_A}||_1 \nonumber \\
& \ge & ||(\hat{P}_0\rho\hat{P}_0)^{T_A}||_1 = ||\rho_\mathrm{trunc}^{T_A}||_1 ,
\end{eqnarray}
or,
\begin{equation} 
E_N(\rho) \ge E_N(\rho_\mathrm{trunc}) .
\end{equation}
This implies that the calculation with the truncated, unnormalized, density matrix does not overestimate the logarithmic negativity of the original state. In particular, if $\rho$ is separable, the negativity of $\rho_\mathrm{trunc}$ necessarily becomes $0$. Furthermore, the difference between the actual negativity and the truncated negativity would be negligible if the truncation number $N_\mathrm{trunc}$ is taken sufficiently large compared to the mean photon number of a given state.

The density matrix elements in the number-state basis can be calculated using the characteristic functions \cite{Barnett} of Eqs. (\ref{eq:cfon}) and (\ref{eq:cfno}) as \begin{align}
& \rho_{ij,kl} = \bra{i,j} \rho \ket{k,l} \nonumber \\
& = \frac{1}{\pi^2}\iint d^2\xi_1 d^2\xi_2 \bra{i,j} \hat{D}_1^\dagger(\xi_1) \hat{D}_2^\dagger(\xi_2) \chi(\xi_1,\xi_2) \ket{k,l}.
\end{align}
In the following, we present numerical results using three squeezing parameters $s=0.029$, $0.114$ and 0.403 for the initial TMSV under different schemes. 
We note that in the pulsed-regime of squeezed light, the squeezing level $s=0.403~(3.5\mathrm{dB})$ is currently achievable using an optical parametric amplifier \cite{Grangier}.
We consider a thermal photon number $n_\mathrm{th}=10^{-5}$ for a low-temperature reservoir and $n_\mathrm{th}=10^{-1}$ for a high-temperature reservoir as examples. 
When we apply a coherent operation $t\hat{a}+r\hat{a}^\dagger$ to each state, the value of $t$  is numerically optimized to yield the highest degree of entanglement under a given situation. We have checked that the local mean photon number is less than 1 for all considered states thus restricted the truncation number to $N_\mathrm{trunc}=5$. 
Furthermore, an increase of $N_\mathrm{trunc}$ to a higher number did not show any appreciable changes in our results.

\subsection{Result}
\begin{figure}[t]
\centering \includegraphics[clip=true, width=0.95\columnwidth]{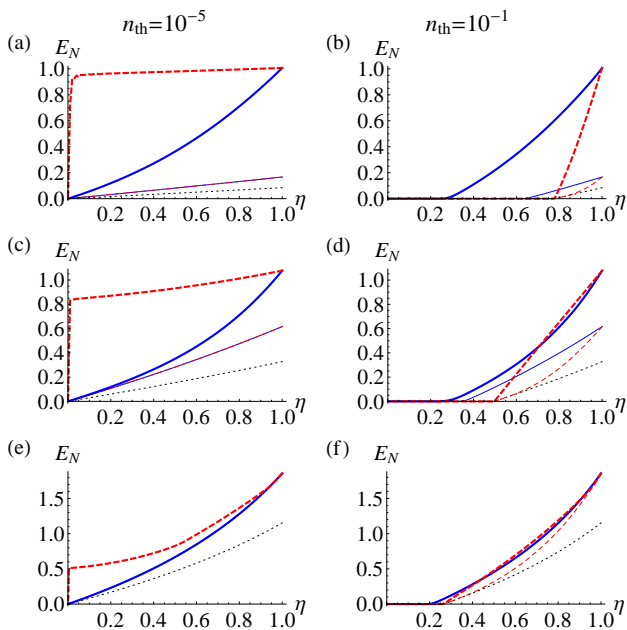}
\caption{\label{fig:negativity}Logarithmic negativity as a function of $\eta=e^{-\Gamma t}$, which characterizes the interaction time with a thermal environment [Eqs. (4) and (5)] under different strategies: no distillation (black dotted), subtraction before noisy channel (thin blue solid), subtraction after noisy channel (thin red dashed), coherent operation before noisy channel (thick blue solid) and coherent operation after noisy channel (thick red dashed). Left (Right) panels represent a low (high)-temperature reservoir with $n_\mathrm{th}=10^{-5}$ ($10^{-1}$). 
Squeezing parameters are given by (a), (b) $s=0.029$, (c), (d) $ s=0.114$ and (e), (f) $s=0.403$. Two thin curves that represent the photon subtraction before and after the noisy channel are overlapped in (a), (c), and (e).}
\end{figure}
In Fig. \ref{fig:negativity}, we plot the logarithmic negativity as a function of $\eta=e^{-\Gamma t}$ that represents the interaction time with a thermal reservoir [Eqs. (4) and (5)]. 
When $\eta=1$, i.e. no reservoir-interaction, it is known that the coherent operation (thick curves) can enhance the entanglement more effectively than the mere photon subtraction (thin curves), which is remarkable particularly in the low-squeezing regime \cite{PhysRevA.84.012302}. This is also true for any value of $\eta <1$ as shown in Fig. \ref{fig:negativity}.
As the environmental interaction becomes longer, i.e. $\eta$ decreases below 1, 
we see that the two strategies applying the coherent operation before and after the noisy channel, respectively, provide an advantage in different temperature regimes. 

When the thermal photon number of the reservoir is very small (left panels), the coherent operation after the noisy channel (dashed curves) generally makes the output entanglement higher. Even for a very low $\eta$, the output entanglement can maintain a rather high value by the coherent operation, particularly for an initially weakly squeezed state [Fig. \ref{fig:negativity} (a)]. Note that the value of the parameter $t$ in the coherent operation $t\hat{a}+r\hat{a}^\dagger$ is optimized for each state case by case in order to maximize the output entanglement in all plots.

The coherent operation or the photon subtraction transforms a Gaussian state to a non-Gaussian state. Therefore, our result implies that keeping Gaussianity through the noisy channel and applying later the coherent operation is advantageous for an optimal distillation in the low-temperature regime. In contrast, it turns out that the photon subtractions before and after the noisy channel (thin solid and dashed curves) do not make any appreciable difference in the output entanglement. Actually, two different strategies yield exactly the same states in the case of vacuum noise ($n_\mathrm{th}=0$) and subtraction operation ($t=1,r=0$). In the case of low temperature, there is difference between two states, however negligible.

On the other hand, when the thermal photon number of the reservoir increases (right panels), the operation before the noisy channel (solid curves) makes the output entanglement higher. This is closely related to the fact that a non-Gaussian entanglement can be more robust than Gaussian entanglement under a high-temperature reservoir, which was shown in Ref. \cite{PhysRevLett.107.238901}. In particular, it is crucial that entanglement can survive, even though very weak in its strength, through a noisy channel: Once entanglement dies out, there is no way of recovering it by LOCC.

When we apply the distillation operation before a noisy channel, entanglement survives longer, e.g. with a small $\eta\gtrsim0.25$ in Figs. \ref{fig:negativity} (b) and (d). However, when we apply the coherent operation after the noisy channel, entanglement can survive shorter, e.g. $\eta\gtrsim0.78$ in (b) and $\eta\gtrsim0.5$ in (d). The separation time of TMSV, at which entanglement completely disappears under a thermal reservoir, can be calculated by Simon's criterion \cite{PhysRevLett.84.2726}. Its analytic expression is given in \cite{PhysRevLett.105.100503} by
\begin{equation}
t_\mathrm{sep}=\frac{1}{\Gamma}\log\left(1+\frac{1-e^{-2s}}{2 n_\mathrm{th}}\right) .
\end{equation}
As $n_\mathrm{th}$ increases, the separation time of TMSV becomes shorter, that is, Gaussian entanglement becomes more fragile under a noisy channel. If we perform distillation before the noisy channel, the resulting non-Gaussian entanglement can survive longer with a smaller $\eta$. Non-Gaussian operations not only increase the degree of entanglement but also make entangled states more robust under noisy channels. We also note that the use of the coherent operation (thick curves) significantly enhances the performance in distillation over the photon subtractions (thin curves). As the squeezing of the initial state increases, however, we see that the difference in the distillation performance between the two strategies becomes smaller [Fig. \ref{fig:negativity} (f)].

\begin{figure}[t]
\centering \includegraphics[clip=true, width=0.6\columnwidth]{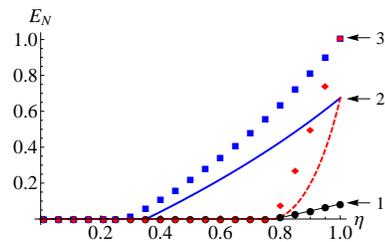}
\caption{\label{fig:cmvsnum}Logarithmic negativity as a function of $\eta=e^{-\Gamma t}$, calculated in two different ways, estimation with covariance matrix (curves) and numerical calculation (symbols: circle, square, diamond), for three different states: no distillation (black thin curve and circle), coherent operation before noisy channel (blue thick curve and square), and coherent operation after noisy channel (red dashed curve and diamond). Squeezing parameter and thermal photon number are $s=0.029$ and $n_\mathrm{th}=10^{-1}$, i.e., the cases of Fig. \ref{fig:negativity} (b).}
\end{figure}

In order to better understand the advantage of non-Gaussian operation applied before the noisy channel, we analyze the logarithmic negativity in more detail for the states appearing in Fig. 2 (b) as examples. In particular, we show in Fig. \ref{fig:cmvsnum} both the degree of Gaussian-type entanglement (curves) that is obtained by the covariance matrix of a given state [Eq. (10)] and the degree of total negativity (symbols) for each state. 

Gaussian entanglement is very robust under a vacuum noise (or a very weak thermal noise) so that it can survive the noise channel very long. Therefore it is deemed a good strategy for the purpose of entanglement distillation that one applies a probabilistic operation after the noise channel. On the other hand, when the strength of thermal noise becomes rather significant, the Gaussian entanglement dies out at a certain value of interaction time [Eq. (15)]. For example, the initial Gaussian entanglement disappears at $\eta\sim0.8$ in Fig. 2 (b). In this case, whatever operation is made on the evolved state after the noisy channel, the entanglement distillation becomes impossible for $\eta<0.8$: Once entanglement dies out, distillation is impossible. Therefore one must perform a certain operation before the state undergoes the noisy channel.

As can be seen from Fig. 3, the non-Gaussian operation (coherent operation) applied before the noise channel ($\eta=1$) increases both the Gaussian-type entanglement (arrow 2) and the total entanglement (arrow 3) from the initial Gaussian entanglement (arrow 1). With the increased entanglement due to the operation, the state becomes more robust than without operation. For example, the enhanced Gaussian-type entanglement (blue thick curve) alone evolves more robustly than the initial two-mode squeezed state (black thin curve) because it has more entanglement (energy) that can resist the noise. Furthermore, due to the non-Gaussian operation performed on the initial state, there also exists a non-Gaussian type entanglement (roughly speaking, the difference between squares and thick solid curve) which can also contribute to the surviving total entanglement. Note that even after the Gaussian-type entanglement disappears at $\eta\sim0.35$, the non-Gaussian entanglement survives until $\eta\sim0.3$.

In a nutshell, at a significant level of thermal noise, the effectiveness of non-Gaussian operation before the noisy channel may be attributed to both the enhanced level of total entanglement (Gaussian and non-Gaussian type) by the operation and the robustness of non-Gaussian entanglement in a certain parameter regime. However, such an effect becomes less remarkable when the initial entanglement is rather high like the case of Fig. 2 (f).

\subsection{Success Probability}
\begin{figure}[t]
\centering \includegraphics[clip=true, width=0.95\columnwidth]{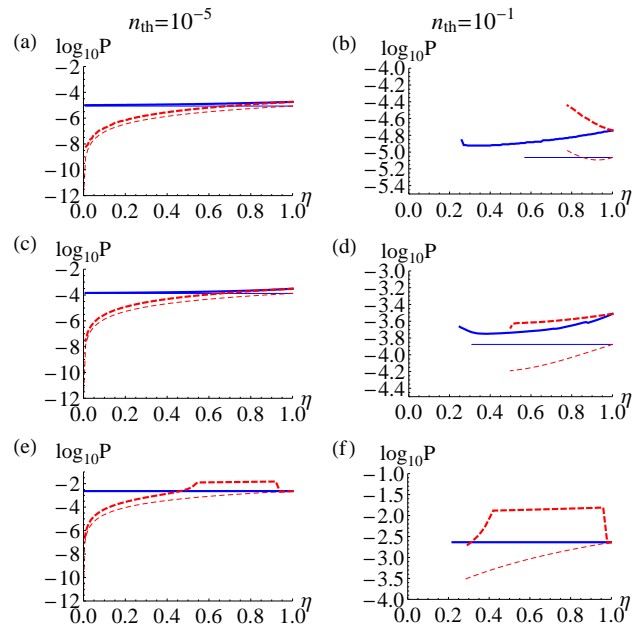}
\caption{\label{fig:probability}Distillation probability (in log scale) as a function of $\eta=e^{-\Gamma t}$. 
Legends and parameters are the same as those in Fig. \ref{fig:negativity}. The curves are missing in the region where distillation is not possible at all, that is, negativity is $0$ before operations.}
\end{figure}

From a practical perspective, it is important to consider the success rate for distillation operation as well as the degree of output entanglement. The success probability of the coherent operation is equal to the probability of detecting a trigger photon, after erasing the which-path information on whether it is from the photon subtraction or addition, for each state. (See the proposed experimental scheme in \cite{PhysRevA.82.053812}.)  The success probability is shown for each case in Fig. \ref{fig:probability}.
We see that the distillation with a coherent operation makes both of the ouput entanglement and the success probability higher than a mere photon subtraction. 
The success rate overall becomes an order of magnitude higher as the initial squeezing increases from $s=0.029$, $s=0.114$ to $s=0.403$. 

In the low-temperature regime, as shown before, we can obtain the output state with high negativity by applying a coherent superposition operation after sending the state through even a very lossy channel (dashed curves), e.g. $\eta\lesssim0.1$. However, the success probability is very low ($\lesssim10^{-6}\sim10^{-4}$) for $\eta=0.1$ and it generally increases with $\eta$. On the other hand, if we apply the coherent operation before the noisy channel (solid curves), we have almost a flat success probability, $\sim10^{-5}$ in (a),$\sim10^{-3.8}$ in (c), and $\sim10^{-2.6}$ in (e) regardless of $\eta$. 

In the high-temperature regime, the success probability for distillation after the noisy channel is higher than that for distillation before the channel. Especially, at $\eta\gtrsim0.7$ in (d), both negativity and probability are higher when we apply operation after sending the state through the channel. However, for small $\eta$ where distillation after the channel is impossible due to the vanishing negativity, we must apply the operations before sending the state through the channel.

Note that the coherent operation is optimized for each case in order to maximize the degree of output entanglement, not the success probability. The success probability is plotted just according to the optimized coherent operation obtained that way, which can affect the behavior of the success probability with respect to $\eta$. 
For the case of operation applied before the noise (blue solid curves), the success probability by its definition is determined before the state evolves under the noise. Thus, the case of photon subtraction (blue thin curves) must have a flat distribution with $\eta$. In cases of Fig. 4 (e) and (f), where the degree of initial squeezing is rather high, it turns out that the optimized coherent operation is also very close to the photon subtraction regardless of $\eta$, thus it also shows a flat distribution.
On the other hand, for the case of operation applied after the noise (red dashed curves), the optimized operation varies with $\eta$. For a rather high (small) $\eta$, the optimized operation approaches the photon addition (subtraction) and the change of the success probability with $\eta$ actually looks more prominent in linear scale than in log scale shown in the figures.

\section{\label{sec:fidelity}Teleportation Fidelity}
The logarithmic negativity studied in the previous section is a measure to quantitatively characterize the degree of entanglement. On the other hand, one may wonder how useful the distributed entangled state can be for informational tasks. We here investigate the usefulness of output entangled state by looking into the CV quantum teleportation \cite{PhysRevLett.80.869}. Specifically, we investigate the teleportation fidelity, an operational measure of entanglement, to show how faithfully a given entangled resource can accomplish the quantum teleportation \cite{PhysRevLett.80.869}. 

It is known that the fidelity between input and output states, averaged over all input coherent states, cannot exceed $1/2$ without entangled resource \cite{PhysRevA.64.022321,Nha}. 
With an entangled resource whose characteristic function is given by $\chi(\xi_1,\xi_2)$, the input-output relation can be written as $ \chi_\mathrm{out}(\xi) = \chi_\mathrm{in}(\xi) \chi(\xi^*,\xi) $ \cite{PhysRevA.74.042306}. The fidelity is then given by
\begin{eqnarray}
F & = & \frac{1}{\pi} \int \mathrm{d}^2\xi \chi_\mathrm{out}(\xi) \chi_\mathrm{in}(-\xi) \nonumber \\
& = & \frac{1}{\pi} \int \mathrm{d}^2\xi \chi(\xi^*,\xi) \chi_\mathrm{in}(\xi) \chi_\mathrm{in}(-\xi) \nonumber \\
& = & \frac{1}{\pi} \int \mathrm{d}^2\xi \chi(\xi^*,\xi) e^{-|\xi|^2}
\end{eqnarray}
for a coherent-state input $ \chi_\mathrm{in}(\xi) = e^{-|\xi|^2/2} e^{\xi\alpha^*-\alpha^*\xi} $.

\begin{figure}[t]
\centering \includegraphics[clip=true, width=0.95\columnwidth]{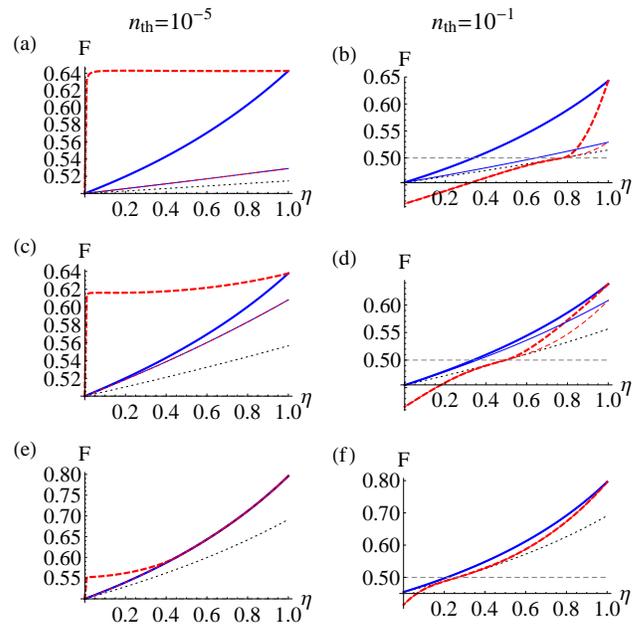}
\caption{\label{fig:fidelity}Teleportation fidelity as a function of transmitivity $\eta$. Legends and parameters are the same as those in Fig. \ref{fig:negativity}. The gray dashed line at $F=0.5$ represents the maximal fidelity with separable resources.}
\end{figure}

Using Eq. (15), we numerically calculate the teleportation fidelity as a function of $\eta=e^{-\Gamma t}$ under two different strategies, which is shown in Fig. \ref{fig:fidelity}. 
Overall, the trend of the fidelity is very similar to that of the logarithmic negativity shown in Fig. \ref{fig:negativity}.
The teleportation fidelity can be high in the parameter region where the negativity is high, but this is not always the case. For example, in Fig. \ref{fig:negativity}(d), we find a crossover between two curves that represent the coherent operation before (solid curve) and after (dashed curve) the noisy channel, while there is no crossover in Fig. \ref{fig:fidelity}(d). This implies that higher entanglement (negativity) does not always provide a more faithful teleportation \cite{Nha}. 
From Fig. \ref{fig:fidelity}, we have a rather clear-cut conclusion about which strategy can yield a better performance for quantum teleportation. In the low (high)-temperature regime, the coherent operation must be applied after (before) sending the TMSV through a thermal reservoir to obtain a higher teleportation fidelity.

\section{\label{sec:conclusion}Conclusion and Remarks}

We have studied the entanglement distillation for CV states by a coherent superposition operation $t\hat{a}+r\hat{a}^\dagger$ when the TMSV decoheres under a thermal noisy channel. 
We showed that a coherent operation still provides a greater advantage than a photon subtraction method in the presence of noise. 
In particular, we considered two different strategies for distillation: a coherent operation applied before and after the noisy channel. 
We investigated two entanglement characteristics, logarithmic negativity and teleportation fidelity, to compare the performances of two strategies. 
In the low-temperature regime, where separation time of TMSV is rather long, the entanglement and the teleportation fidelity for the output state can be higher if we apply the operation after the noisy channel. The success probability for distillation decreases with the channel transmissivity $\eta=e^{-\Gamma t}$. On the other hand, in the high-temperature regime, the entanglement of TMSV decays rapidly and no distillable entanglement remains under a low transmissivity. It turns out to be better to apply operation before sending the TMSV through the channel so that the state becomes non-Gaussian with higher entanglement that can be more robust under a high-temperature noisy channel. In the limit of a large initial squeezing of the  TMSV state, both strategies give almost the same degree of performance in distillation.

In a realistic situation, the thermal photon number in environment can be very small. For example, the average thermal photon for the wavelength $\lambda=1064\mathrm{nm}$ is $n_\mathrm{th}=2.61\times10^{-20}$ at room temperature. In this case, the behavior of entanglement is similar to what we have shown with $n_\mathrm{th}=10^{-5}$. However, the average thermal photon number becomes as large as $n_\mathrm{th}=10^{-1}$ in the infrared regime ($\lambda=20\mu\mathrm{m}$) at room temperature and it becomes much larger in the microwave regime. We may also consider other kinds of noisy attenuator environment which can be modeled as a beamsplitter interaction with thermal photons. Therefore, a large photon number like $n_\mathrm{th}=10^{-1}$ may be considered as realistic.

Finally, let us briefly address the applicability of distillation with a coherent superposition operation for a long-distance quantum communication. 
In the low-temperature regime, we can obtain a highly entangled state when the coherent operation is successfully applied after the noisy channel. Thus the coherent operation alone can make a good distillation protocol to distribute entanglement between two distant nodes. However, in the high-temperature regime, transmission distance is limited and the degree of distributed entanglement is low. For a long-distance quantum communication, we may need to combine the coherent operation with other protocols such as quantum repeater and multiple-copy distillation. Recently, CV distillation protocols have been studied with single photon subtraction and two-copy mixing \cite{PhysRevA.67.062320,PhysRevLett.108.060502}. It may be worth further studying whether combining the coherent operation and two-copy mixing can be an efficient distillation tool.

\end{document}